\documentclass[
    ,final            % use final for the camera ready runs
  ]
  {aipproc}

\usepackage{amssymb}
\layoutstyle{8x11single}

\begin{document}

\title{Tomographic entropic inequalities in the probability
representation of quantum mechanics}

\classification{42.50.Ct, 42.50.Ex, 02.50.Cw}
\keywords      {Entropy, uncertainty relations,
modified tomogram, universal inequality for the wave function.}

\author{Margarita A. Man'ko and Vladimir I. Man'ko}{
  address={P. N. Lebedev Physical Institute,
  Leninskii Prospect 53, Moscow 119991, Russia}
}

\begin{abstract}
A review of the tomographic-probability  representation of classical
and quantum states is presented. The tomographic entropies and entropic
uncertainty relations are discussed in connection with ambiguities in
the interpretation of the state tomograms which are considered either
as a set of the probability distributions of random variables depending
on extra parameters or as a single joint probability distribution of
these random variables and random parameters with specific properties
of the marginals. Examples of optical tomograms of photon states, symplectic
tomograms, and unitary spin tomograms of qudits are given. A new universal
integral inequality for generic wave function is obtained on the base of
tomographic entropic uncertainty relations.
\end{abstract}

\maketitle

%%%%%%%%%%%%%%%%%%%%%%%%%%%%%%%%%%%%%%%%%%%%
%% MAINMATTER
%%%%%%%%%%%%%%%%%%%%%%%%%%%%%%%%%%%%%%%%%%%%

\section{Introduction}
In a conventional approach, the states of quantum systems are described
by the wave function~\cite{Schrod26}, the density matrix~\cite{Landau,vonNeum}
or a vector in the Hilbert space~\cite{Diracbook}.
In the probability representation of quantum mechanics~\cite{ManciniPLA}, the
states of quantum systems are described by the tomographic-probability
distributions or tomograms (see, for example, the review~\cite{IbortPS}).
It was shown~\cite{OlgaJRLR97,MendesPhysicaD} that the states of classical
systems can also be described by the tomograms used as alternatives to the
probability densities on the phase space.

Since any probability distribution provides such informational and statistical
characteristics as Shannon~\cite{Shannon} and R\`{e}nyi~\cite{Renyi} entropies,
the corresponding tomographic entropies were introduced~\cite{OlgaJRLR97,Rita-entropy},
and the properties of such entropies were discussed (see, for example,
\cite{RitaFP}). Among these properties, there are entropic uncertainty relations
known for the probability distributions associated with the wave functions (see,
for example, \cite{Hirschman,183}) and density matrices (see, for example, \cite{Birula}).
Recently, the tomographic entropic uncertainty relations obtained in \cite{ActaHung06,RitaFP}
have been confirmed experimentally~\cite{BellFilipPRA} with an accuracy of a few percents.

The tomograms for both the continuous photon quadratures~\cite{ManciniPLA} and the discrete spin
variables~\cite{OlgaJETP,DodPLA} have some specific properties. The tomograms are the distribution functions
of random variables, and these functions depend on extra parameters. For photon
optical tomograms~\cite{BerBer,VogRis}, the probability distribution $w(X,\theta)$
depends on a random homodyne real quadrature $-\infty<X<\infty$ and the local
oscillator phase $0\leq\theta\leq 2\pi$ considered as a control parameter.
For unitary spin tomogram $w(m,u)$~\cite{Sud}, the function depends on a random discrete
variable $-j\leq m\leq j$, which is the spin projection on the quantization axes
obtained by the action of a unitary rotation matrix $u$ on the initial $z$ axis.

The aim of this paper is to demonstrate some ambiguities in such interpretation.
We present the other possibility to interpret the tomograms as joint probability
distributions of two random variables --- considering the extra parameters as random
variables. For optical tomograms, such interpretation was discussed in \cite{Ital-from-Bel,BellFilipPRA}.
Such interpretation gives some new clarification of the properties of classical
and quantum tomograms and provides with possibilities of the generalization of
known tomograms by introducing many other joint probability distributions containing
the same information on quantum (also classical) states as the initial optical or
symplectic tomograms for the continuous homodyne quadratures and discrete spin variables
for spin tomograms. The other goal is to reformulate some entropic uncertainty relations
for pure states as a new integral inequality for the wave functions.

\section{Entropies in the probability theory}
In the information-theory context, entropy is related to an arbitrary
probability-distribution function. We remind the notion of Shannon entropy~\cite{Shannon}.
Given the probability distribution $P(n)$, where $n$ is a discrete random variable, i.e.,
$~P(n)\geq 0$,
and the normalization condition holds
$~\sum_nP(n)=1$,
one has, by definition, the Shannon entropy
\begin{equation}\label{A3}
S=-\sum_nP(n)\ln P(n)=-\langle\ln P(n)\rangle.
\end{equation}
There exist other kinds of entropies depending on extra parameter $q$
and associated to the probability distribution $P(n)$, for example,
R\'enyi entropy~\cite{Renyi}
\begin{equation}\label{Aa}
R(q)=\frac{1}{1-q}\,\ln\left[\sum_n\Big(P(n)\Big)^q\right],\qquad q>0.
\end{equation}
In the limit $q\to 1$, one has the equality of the R\'{e}nyi entropy to the Shannon entropy
$~R(1)=S.$
The Shannon entropy $S$ is a number. The R\'{e}nyi entropy $R(q)$ is the function of
a parameter $q$ and, due to this, it contains more information on details of the
probability distribution $P(n)$ including the value of Shannon entropy.

For distribution functions $P(x)$, in the case of continuous random variable $x$, one
has the same definition with the replacements
%\begin{equation}\label{Dd}
$~n\to x~~$ and $~\sum_n\to\displaystyle{\int} dx.$
%\end{equation}
For the case of quantum-system states with density matrix $\rho$, one has an analog
of the Shannon entropy, which is called the von Neumann entropy, given by formula
with taking the trace
\begin{equation}\label{Ee}
S_{\rm vN}=-\mbox{Tr}\left(\rho\ln\rho\right),
\end{equation}
and quantum R\'enyi entropy
\begin{equation}\label{Ff}
R_\rho(q)=\frac{1}{1-q}\,\ln\mbox{Tr}\left(\rho^q\right).
\end{equation}

The above formulas generalize the definition of entropies in the classical domain.
In the limit $q\to 1$, for quantum system entropies, one has an equality analogous to
the previous equality $~R(1)=S$,
i.e., one has the equality of the quantum R\'eniy entropy to the von Neumann entropy
$~R_\rho(1)=S_{\rm vN}$.

Below we apply the introduced definitions of entropy in all cases of
probability distributions and density matrices to tomograms, since the tomograms
themselves are fair probability distributions for both classical and quantum systems.

\section{Specific probability distributions and their marginals}
Now we consider some special probability distributions $P(a,b)$ of a random variable $a$,
which depends also on extra parameters $b$, where $a$ and $b$ denote some sets of variables,
both continuous and discrete ones. The functions $P(a,b)$ are nonnegative $~P(a,b)\geq 0$
and normalized $~\sum_aP(a,b)=1$ for arbitrary values of the parameters $b$.
This property of the function $P(a,b)$ is called ``no signaling.''
The functions $P(a,b)$ can be obtained from a joint probability distribution $f(a,b)\geq 0$
of two random variables satisfying the normalization condition $~\sum_{a,b}f(a,b)=1$.
In fact, one can construct the marginal $~K(b)=\sum_{a}f(a,b)$ and define the function $P(a,b)$
by the relation $~P(a,b)=f(a,b)K^{-1}(b)$. In this case, the function $P(a,b)$ is called
the conditional probability distribution of random variable $a$ provided the output of
the second event is known.

A simple example of such a function is the case where $a=1,2$, $b=1,2$, $P(1,1)=x$, $P(1,2)=1-x$,
$P(2,1)=y$, $P(2,2)=1-y$, and $0\leq x,y\leq 1$. In this example, the function $P(a,b)$
corresponds to the set of two different probability distributions $P(a,1)$, such that $P(1,1)=x$
and $P(2,1)=1-x$, and $P(a,2)$, such that $P(1,2)=y$ and $P(2,2)=1-y$.
The structure of this function provides the possibility to consider it as a single joint
probability distribution ${\cal P}(a,b)$ of two random variables determined by the formula
\begin{equation}\label{PM1}
{\cal P} (a,b)=P(a,b)/2.
\end{equation}
In fact, $~{\cal P}(a,b)\geq 0$ and $~\sum_{a,b}{\cal P}(a,b)=1$. This joint probability
distribution has two marginals, namely,
\begin{equation}\label{PM2}
\Pi_1(a)=\sum_b{\cal P}(a,b),\qquad \Pi_2(b)=\sum_a{\cal P}(a,b).
\end{equation}
One can see that
\begin{eqnarray}
\Pi_1(1)=(x+y)/2,&\qquad&\Pi_1(2)=1-(x/2)-(y/2),\label{PM3}\\
\Pi_2(1)=1/2,&\qquad&\Pi_2(2)=1/2.\label{PM4}
\end{eqnarray}
If one considers the function $\Pi_2(b)$ as a probability distribution, it corresponds to
maximum chaotic behavior of the random variable $b$ with maximum Shannon entropy $S=\ln 2$.

The generic joint probability distribution is determined by three nonnegative parameters
$p_1$, $p_2$, and $p_3$ as follows:
\begin{equation}\label{PM5}
\Pi(1,1)=p_1,\qquad\Pi(1,2)=p_2,\qquad\Pi(2,1)=p_3,\qquad\Pi(2,2)=1-p_1-p_2-p_3.
\end{equation}
The above distributions, for which one of the marginals coincides with the distribution
with maximum entropy, is determined by two parameters. Thus, we can see that the joint
probability distributions with such specific properties of the marginals belong to a
subdomain in the simplex corresponding to a set of generic probability distributions
of two random variables. It is clear that an analogous situation can be found for
the other functions and the other domains of the variables $a$ and $b$. Also,
for the function $P(a,b)$, a new joint probability distribution $W(a,b)$ can be
constructed using an arbitrary probability distribution $w(b)$ of a random variable $b$
as follows: $~W(a,b)=P(a,b)w(b)$.

In fact, $~W(a,b)\geq 0$ and $~\sum_{a,b}W(a,b)=1$, since $~\sum_{b}w(b)=1$, as well as
$~\sum_aP(a,b)=1$. The permutation symmetry $a\rightleftarrows b$ can take place.

The other examples of optical, symplectic and spin tomograms, which have analogous
``no signaling'' properties,  are given in the next sections.

\section{Optical and symplectic tomograms}

\subsection{Tomographic-probability distributions of classical particles}

Given the probability density $f(q,p)$ on the phase space. The function $f(q,p)$,
due to the physical meaning of the probability distribution,
is nonnegative and normalized
$~\displaystyle{\int} f(q,p)\,dq\,dp=1$.
Let us calculate the marginal probability density of the particle's
position $X$ in a rotated reference frame on the phase space with new rotated axes.
One has the expression for position $X$ in rotated reference frame as follows:
\begin{equation}\label{1.2}
X=q\cos\theta+p\sin\theta,
\end{equation}
where $\theta$ is the rotation angle.
One can see that for $\theta=0$, $X=q$ and for $\theta=\pi/2$, $X=p$.

The marginal probability density $w(X,\theta)$ (called optical tomogram in quantum optics,
but we will use this name also in classical statistical mechanics) reads
\begin{equation}\label{1.3}
w(X,\theta)=\langle\delta\left(X-q\cos\theta-p\sin\theta\right)\rangle=
\int f(q,p)\delta\left(X-q\cos\theta-p\sin\theta\right)\,dq\,dp.
\end{equation}

Now we introduce another tomogram (related to the optical tomogram) accompanying the rotation
of reference frame in the phase space by scaling the position and momentum before the rotation.
Namely, we consider the marginal probability density denoted as $M(X,\mu,\nu)$ of the
particle's position $X$ in a reference frame on the phase space, which first was rescaled
and then was rotated; it reads
\begin{eqnarray}\label{1.4}
M(X,\mu,\nu)=\langle\delta\left(X-\mu q-\nu p\right)\rangle=
\int f(q,p)\delta\left(X-\mu q-\nu p\right)\,dq\,dp,\nonumber\\
\end{eqnarray}
where $\mu$ and $\nu$ could be arbitrary real numbers.
The probability distribution $M(X,\mu,\nu)$ is called
symplectic tomogram of the classical particle's state.
It is normalized $~\displaystyle{\int} M(X,\mu,\nu)\,dX=1$,
due to the property of delta-function
$~\displaystyle{\int}\delta(X-\mu q-\nu p)\,dX=1$
and the normalization of the distribution $f(q,p)$
on the phase space.
Due to the homogeneity of the Dirac delta-function, i.e.,
$\delta(\lambda x)=|\lambda|^{-1}\delta (x)$, the symplectic
tomogram is also the homogeneous function,
$M(\lambda X,\lambda\mu,\lambda\nu)=|\lambda|^{-1}M(X,\mu,\nu)$.
Thus, one has the connection between the optical and symplectic
tomograms, due to the homogeneity property, i.e.,
\begin{eqnarray}
w(X,\theta)&=&M(X,\cos\theta,\sin\theta),\label{1.5}\\
M(X,\mu,\nu)&=&\frac{1}{\sqrt{\mu^2+\nu^2}}\,w\left(\frac{X}{\sqrt{\mu^2+\nu^2}}\,,
\mbox{tan}^{-1}\frac{\nu}{\mu}\right).\label{1.6}
\end{eqnarray}
Formulae for the optical tomogram, given by its definition, namely,
(\ref{1.3}),
and symplectic tomogram, given by its definition (\ref{1.4}),
turn out to coincide with the well-known integral Radon transform of the function
of two variables $f(q,p)$, which has the inverse.
The inverse reads
\begin{equation}\label{1.7}
f(q,p)=\frac{1}{4\,\pi^2}\int M(X,\mu,\nu)e^{i(X-\mu q-\nu p)}dX\,d\mu\,d\nu\geq 0.
\end{equation}
The probability distribution $M(X,\mu,\nu)$ can be used to calculate momenta of the
random variables $q$ and $p$.
In fact, due to the physical meaning of the marginal probability distribution
$M(X,\mu,\nu)$, one has
\begin{equation}
\langle q^n\rangle =\int M(X,1,0)X^n\,dX,\qquad %label{1.8}\\
\langle p^n\rangle =\int M(X,0,1)X^n\,dX.\label{1.9}
\end{equation}

\subsection{Tomographic-probability distributions of quantum particles}
The quantum particle's state can be described by the tomogram obtained using
the formula for classical tomogram with averaging the delta-function~(\ref{1.3})
So, we start from this formula keeping only the form with averaging
$w(X,\theta)=\langle\delta\left(X-q\cos\theta-p\sin\theta\right)\rangle,$
but with the following replacement in this form the numbers $q$ and $p$
by the corresponding operators, i.e., the position is replaced by the position
operator $q\to\hat q$, and the momentum is replaced by the momentum operator
$p\to\hat p$. Also the classical averaging has to be replaced with averaging
by means of the quantum-state density operator $\hat\rho.$

For photon states, the photon quadrature components play the role of position $q$
and momentum $p$. Then, for the photon quantum state in quantum optics,
the optical tomogram is defined as
\begin{equation}\label{2.1}
w(X,\theta)=\langle\delta\left(X-\hat q\cos\theta-\hat p\sin\theta\right)\rangle.
\end{equation}
The average means that we replaced the probability distribution $f(q,p)$ in the definition
of classical optical tomogram by the density operator, i.e.,
$f(q,p)\to\hat\rho$ and applied the formula for average of the operator $\hat A$ of the form
$\langle \hat A\rangle=\mbox{Tr}\left(\hat\rho\hat A\right).$

The definition of optical tomogram~(\ref{2.1})
can be done in a more known form (see, \cite{BerBer,VogRis})
which uses the Wigner function $W(q,p)$ of the photon
quantum state
$$w(X,\theta)=\int W(q,p)\delta(X-q\cos\theta-p\sin\theta)\frac{dq\,dp}{2\pi}\,.$$
One can see that the optical tomogram of classical particle is given by the same
formula with replacement $~W(q,p)/2\pi\to f(q,p).$
The above optical tomogram of the photon quantum state is measured by homodyne
detector~\cite{Raymer}.

The symplectic tomogram of the quantum state is given by the classical formula (\ref{1.4})
with the same replacements $~q\to\hat q$ and $~p\to\hat p$, i.e.,
\begin{equation}\label{2.2}
M(X,\mu,\nu)=\langle\delta\left(X-\mu\hat q-\nu\hat p\right)\rangle.
\end{equation}
The quantum tomogram $M(X,\mu,\nu)$
determines the density operator $\hat\rho$ by the formula analogous
to the classical formula for reconstructing the probability distribution $f(q,p)$ on the phase space
but with the replacement $f(q,p)\to\hat\rho$, $q\to\hat q$, $p\to\hat p$, and $1/4\pi^2\to1/2\pi$,
i.e.,
\begin{equation}\label{15}
\hat\rho=\frac{1}{2\pi}\int M(X,\mu,\nu)e^{i\left(X-\mu\hat q-\nu\hat p\right)}\,dX\,d\mu\,d\nu.
\end{equation}
One can see that inverse Radon transform (\ref{1.7})
for the classical symplectic tomogram $M(X,\mu,\nu)$
coincides with its Fourier transform.
Reconstruction formula (\ref{15}) for the quantum density operator $\hat\rho$ has the form
of ``quantized" Fourier transform of the quantum symplectic tomogram $M(X,\mu,\nu)$.

We summarize the notion of classical and quantum states in terms of tomograms $M(X,\mu,\nu)$
in the tomographic-probability representation as follows.
\begin{itemize}
\item
The states in both classical and quantum mechanics can be associated with nonnegative
normalized homogeneous probability
distributions $M(X,\mu,\nu)$ (tomograms) depending on a random variable $X$ and real
parameters $\mu$ and $\nu$.
\item
The quantum optical and symplectic tomograms satisfy the same formulae (\ref{1.5})
and (\ref{1.6}) like the classical tomograms.
This means that measuring the quantum optical tomogram $w(X,\theta)$ by homodyne detector
implies measuring the symplectic tomogram. Namely in homodyne experiments one can study
optical tomograms and entropic inequalities which distinguish the classical and quantum
domains.
\end{itemize}

\section{Modified optical, symplectic, and spin tomograms}
The optical and symplectic tomograms introduced have a form of the function $P(a,b)$
discussed in the previous sections. In fact, for the optical tomogram the variable $a$
is the homodyne quadrature $X$, and the variable $b$ is the local oscillator phase $\theta$.

So we can introduce a modified optical tomogram
\begin{equation}\label{2.2a}
W(X,\theta)=w(X,\theta)R(\theta),
\end{equation}
where $R(\theta)\geq 0$ and $\displaystyle{\int}_0^{2\pi}R(\theta)\,d\theta=1$.
Thus, $R(\theta)$ is an arbitrary probability density on a circle; for example, we can use
$R(\theta)=(2\pi)^{-1}$.

For symplectic tomogram, one can provide a modification of the form
\begin{equation}\label{2.3a}
\widetilde M(X,\mu,\nu)=M(X,\mu,\nu)R(\mu,\nu),
\end{equation}
where $R(\mu,\nu)\geq 0$ and $\displaystyle{\int\!\!\!\!\int} R(\mu,\nu)\,d\mu\,d\nu=1$.
Thus, $R(\mu,\nu)$ can be taken as an arbitrary probability density on the plane $(\mu,\nu)$.
For example, we can use the Gaussian distribution function.

Summarizing, for both the classical and quantum cases, we have introduced a modified optical
tomogram which is the joint probability distribution of the homodyne quadrature component
and the local oscillator phase.
For symplectic tomogram of the classical state, one can introduce a modified version
of the Gaussian form
\begin{equation}\label{2.4a}
\widetilde M_G(X,\mu,\nu)=\frac{1}{\pi}\int f(q,p)\left[\delta(X-\mu q-\nu p)
\exp\left(-\mu^2-\nu^2\right)\right]dq\,dp.
\end{equation}
The inversion formula reads
\begin{equation}\label{2.5a}
f(q,p)= \frac{1}{4\pi}\int \widetilde M_G(X,\mu,\nu)\exp
\left[\mu^2+\nu^2+i\left(X-\mu q-\nu p\right)\right]dX\,d\mu\,d\nu.
\end{equation}
For the quantum case, the modified optical tomogram reads
\begin{equation}\label{O1}
W(X,\theta)=\langle\delta\left(X-\hat q\cos\theta-\hat p\sin\theta\right)R(\theta)\rangle.
\end{equation}
The modified symplectic tomogram of quantum state can be defined using
the Gaussian factor as follows:
\begin{equation}\label{O2}
\widetilde M(X,\mu,\nu)=\frac{1}{\pi}
\langle\delta\left(X-\mu\hat q-\nu\hat p\right)\exp\left(-\mu^2-\nu^2\right)\rangle,
\end{equation}
and the inverse of (\ref{O2}) is
\begin{equation}\label{O3}
\hat\rho=\frac{1}{2}\int\widetilde M(X,\mu,\nu)\exp\left[\mu^2+\nu^2
+i\left(X-\mu\hat q-\nu\hat p\right)\right]dX\,d\mu\,d\nu.
\end{equation}
One can also make a modification of the same kind of the unitary spin tomogram.
The tomogram $w(m,u)$ reads~\cite{OlgaJETP,DodPLA,ZacSudPLA,OcastaJPA}
\begin{equation}\label{S1}
w(m,u)=\langle m\mid u\hat\rho u^\dagger\mid m\rangle,
\end{equation}
and it is the function of spin projection $-j\leq m\leq j$ and the unitary-group
element $u$.

If the matrix $u$ coincides with the matrix of irreducible representation of
the group $SU(2)$, the tomogram is the function $w(m,\vec n)$ of the spin projection
$m$ depending on the quantization direction $\vec n$. The spin tomograms $w(m,u)$ and
$w(m,\vec n)$ are nonnegative and normalized functions
\begin{equation}\label{S2}
\sum_{m=-j}^jw(m,u)=1,\qquad \sum_{m=-j}^jw(m,\vec n)=1
\end{equation}
for arbitrary directions $\vec n$ and arbitrary unitary matrices $u$. This means
that the tomograms belong to the set of functions $P(a,b)$, which can be related to
functions ${\cal P}(a,b)$ discussed above.
In view of this, one can introduce modified spin tomograms. One of the modifications
reads
\begin{equation}\label{S3}
\widetilde w(m,\vec n)=w(m,\vec n)R(\vec n),
\end{equation}
where $R(\vec n)$ is any probability density on the sphere $S^2$, i.e.,
$R(\vec n)\geq 0$ and the integral over the sphere $~\displaystyle{\int} R(\vec n)\,d\vec n=1$.

The modified unitary spin tomogram reads
\begin{equation}\label{O4}
\widetilde w(m,u)=w(m,u)R(u),
\end{equation}
where $R(u)$ is any probability density on the unitary group, i.e.,
$R(u)\geq o$ and $~\displaystyle{\int} R(u)\,du=1$, with $du$ being the Haar
measure on the group, $~\displaystyle{\int} du=V$,
and $V$ the volume on the unitary group. For example, one can consider a maximum
chaotic distribution $R(u)=1/V$ with the Shannon entropy $S_u=\ln\, V$.

In the case of modified spin tomogram $\widetilde w(m,\vec n)$, we can take
the distribution $R(\vec n)=1/4\pi$ corresponding to the area of the unit-radius
sphere $~\displaystyle{\int} d\vec n=4\pi$. This maximum chaotic distribution
has the Shannon entropy $S_{\vec n}=\ln \,4\pi$.

Thus, we introduced the modified spin tomograms, which are functions of two sets
of random variable corresponding to functions $P(a,b)$, where $a$ is the spin projection
$m$, and $b$ is either a point on the unit sphere $S^2$ parametrized by the unit vector
$\vec n$ or the element of the unitary group. It is worth noting that all other available
tomographic-probability distributions like the photon-number tomograms~\cite{TombesiEPL}
or the center-of-mass tomograms~\cite{ArxipPRA} can also be modified in an analogous way.
One can see that there exists an ambiguity in choosing the tomographic-probability
distributions which can be associated with the states in both classical and quantum
domains. The ambiguity is related to the choice of the probability distribution of random
parameters.

\section{ Modified Tomographic Entropies}
Since the symplectic tomogram is the standard probability
distribution, one can introduce entropy associated with the tomogram of quantum
state~\cite{OlgaJRLR97} or with the tomogram of analytic signal~\cite{Rita-entropy}.
Thus one has entropy as the function of two real variables $\mu$ and $\nu$
\begin{equation}\label{T33}
S(\mu,\nu)=-\int M(X,\mu,\nu)\,\ln M(X,\mu,\nu)\,dX.
\end{equation}
We call this entropy the symplectic entropy.
In view of the homogeneity and normalization conditions for
tomograms, one has the additivity property
\begin{equation}\label{T34}
S(\lambda\mu,\lambda\nu)=S(\mu,\nu)+\ln|\lambda|.
\end{equation}
Also one has the optical tomographic entropy associated with
the optical tomogram $w(X,\theta)$ as
\begin{equation}\label{T34a}
S(\theta)=-\int w(X,\theta)\ln w(X,\theta)\,dX,
\end{equation}
and this entropy depends on local oscillator phase in experiments with
measuring photon homodyne quadrature.

Since we introduced the modified optical and symplectic tomograms, modified
tomographic entropies can be defined.

For symplectic tomogram, modified tomographic entropy reads
\begin{equation}\label{Aa1}
S^{(\rm sym)}=\langle S(\mu,\nu)\rangle+S^{(\rm sym)}_R,
\end{equation}
where
\begin{eqnarray}
\langle S(\mu,\nu)\rangle=\int d\mu\,d\nu\,R(\mu,\nu)S(\mu,\nu),\label{Aa2}\\
S_R=-\int R(\mu,\nu)\ln R(\mu,\nu)\,d\mu\,d\nu.\label{Aa3}
\end{eqnarray}

For optical tomogram, modified tomographic entropy is
\begin{equation}\label{Aa4}
S^{(\rm opt)}=\langle S(\theta)\rangle+S^{(\rm opt)}_R,
\end{equation}
where
\begin{eqnarray}
\langle S(\theta)\rangle=\int_0^{2\pi} d\theta\,S(\theta)R(\theta),\label{Aa5}\\
S_R^{(\rm opt)}=-\int^{2\pi}_0 R(\theta)\ln\left(R(\theta)\right)\,d\theta.\label{Aa3}
\end{eqnarray}

Analogous modified tomographic entropy can be defined for spin tomograms.

The quantum optical tomogram of the pure state is determined by the wave function
as (see, for example, \cite{RitaFP})
\begin{equation}\label{T18}
w(X,\theta)=\left|\int\psi(y)\exp\left[\frac{i}{2}\left(\mbox{cot}\,\theta\,\,
(y^2+X^2)-\frac{2X}{\sin \theta}\,y\right)\right]\frac{dy}{\sqrt{2\pi
i\sin \theta}}\right|^2.
\end{equation}
On the other hand, this tomogram formally equals to
\begin{equation}\label{T21}
w(X,\theta)=|\psi(X,\theta)|^2,
\end{equation}
where the wave function reads
\begin{equation}\label{T22}
\psi(X,\theta)=\frac{1}{\sqrt{2\pi i\sin \theta}}\int
\exp\left[\frac{i}{2}\left(\mbox{cot}\,\theta\,\,
(y^2+X^2)-\frac{2X}{\sin \theta}\,y\right)\right]\psi(y)\,{dy},
\end{equation}
being the fractional Fourier transform of the wave function
$\psi(y)$. This wave function corresponds to the wave function of
a harmonic oscillator with $\hbar=m=\omega=1$ taken at the ``time''
moment $\theta$ provided the wave function at the initial time moment
$\theta=0$ equals to $\psi(y)$.

In view of expressions of tomogram in terms of the wave function
(\ref{T21}) and (\ref{T22}), one has the entropic uncertainty relation in the form
\begin{equation}\label{T46}
S(\theta)+S(\theta+\pi/2)\geq\ln \pi e,
\end{equation}
which is the Hirshman uncertainty relation
\begin{equation}\label{H1}
-\int|\psi(x)|^2\ln|\psi(x)|^2\,dx-\int|\widetilde\psi(p)|^2
\ln|\widetilde\psi(p)|^2\,dp\geq\ln \,\pi e,
\end{equation}
considered in a rotated reference frame on the phase space~\cite{Hirschman,RitaFP},
with $\widetilde\psi(p)$ being the wave function in the momentum representation.
In (\ref{T46}), $S(\theta)$ is the tomographic Shannon entropy associated
with optical tomogram~(\ref{T18}) which is measured by homodyne detector.

One can write the subadditivity and strong subadditivity conditions for modified
spin tomograms. For example, using Eq.~(\ref{S3}), we obtain the subadditivity
condition of the form
\begin{equation}\label{Cc1}
-\sum_{\vec n}\left(\sum_m\widetilde w(m,\vec n)\ln \left[\sum_m\widetilde w(m,\vec n)\right]\right)-
-\sum_m\left(\sum_{\vec n}\widetilde w(m,\vec n)\ln \left[\sum_{\vec n}\widetilde w(m,\vec n)\right]\right)
\geq-\sum_m\sum_{\vec n}\widetilde w(m,\vec n)\ln \widetilde w(m,\vec n),
\end{equation}
where we used several (arbitrary number) different directions $\vec n$ such that
$\sum_{\vec n}R(\vec n)=1$.

For two qudits, the modified tomogram of the state with density matrix $\rho(1,2)$ can be given as
\begin{equation}\label{Cc2}
\widetilde w(m_1,m_2,u)=\langle m_1m_2\mid u\rho(1,2)u^\dagger\mid m_1m_2\rangle R(u),
\end{equation}
where for the distribution $R(u)$ one can take several (arbitrary number) different matrices
$u$ such that $\sum_uR(u)=1$. One has the strong subadditivity condition
\begin{equation}\label{Cc3}
S(1,2)+S(2,3)\geq S(1,2,3)+S(2),
\end{equation}
where $S(1,2)$ and $S(2,3)$ are Shannon entropies for marginal distributions
\begin{equation}\label{Cc4}
\widetilde\Omega(m_1,u)=\sum_{m_1}\widetilde w(m_1,m_2,u)\qquad \mbox{and}\qquad
\widetilde\Omega(m_2,u)=\sum_{m_2}\widetilde w(m_1,m_2,u),
\end{equation}
respectively. The entropy $S(1,2,3)$ is the Shannon entropy for distribution (\ref{Cc2}) and
$S(2)$ is the Shannon entropy for distribution $\widetilde\Omega(u)=\sum_{m_1,m_2}\widetilde w(m_1,m_2,u)$.

Using (\ref{T18}) and integrating (\ref{T46}) over the local oscillator phase
$0\leq\theta\leq 2\pi$, we obtain the inequality
\begin{eqnarray}\label{Ww}
&&-\int_0^{2\pi}\!\!\!\!\!\int_{-\infty}^\infty \frac{d\theta\,dX}{|\sin\theta|}\left|
\int_{-\infty}^\infty\psi(y)\exp\left(\frac{i\cot\theta}{2}y^2-\frac{iXy}{\sin\theta}
\right)dy\right|^2\nonumber\\
&&\times\ln\left[\frac{1}{2\pi|\sin\theta|}\left|
\int_{-\infty}^\infty\psi(z)\exp\left(\frac{i\cot\theta}{2}z^2-\frac{iXz}{\sin\theta}
\right)dz\right|^2\right]\geq 2\pi^2\ln\pi e.
\end{eqnarray}
This universal integral inequality must be fulfilled for an arbitrary wave function $\psi(y)$,
satisfying the normalization condition $\displaystyle{\int}_{-\infty}^\infty|\psi(y)|^2dy=1$.
The entropic inequality in the form of inequality for the zero Fourier component of the function
of $\theta$ in Eq.~(\ref{T46}) was obtained in \cite{CosimoCQG} and in the form of integral
inequality containing the optical tomogram and checked experimentally in \cite{BellFilipPRA}.

Entropic inequality~(\ref{Ww}) is obvious in the tomographic-probability representation
of quantum states but in the standard formulation of quantum mechanics it becomes more
complicated integral inequality for the wave function. This inequality could be related either
to the properties of optical tomograms considered as the function of one random variable $X$
or as the joint probability distribution of random variables $X$ and $\theta$.

\section{Conclusions}
To conclude, we summarize the main results of our work.
\begin{itemize}
\item
We showed that all the available state tomograms can be considered either as the probability
distributions of random variables depending on extra parameters with no signaling properties
or as the joint probability distributions of both sets of variables and the parameters.
\item
We presented possible modifications of optical, symplectic, and spin tomograms.
\item
We studied properties of the wave function $\psi(y)\in L_2$ for the available optical
tomographic entropic inequality associated with the tomograms and obtained the universal
integral inequality for an arbitrary wave function.
\item
We clarified the ambiguity in choosing the tomographic-probability distribution describing
the states in both the classical and quantum domains.
\end{itemize}

\section*{Acknowledgments}
This study was supported by the Russian Foundation for Basic Research
under Projects Nos.~10-02-00312 and 11-02-00456. The authors thank the
Universidad Nacional Autonoma de Mexico and Prof. Octavio Casta\~nos and
the Organizers of the conference ``Beauty in Physics: Theory and Experiment"
(the Hacienda Cocoyoc, Morelos, Mexico, May 14--18, 2012) Profs. Alejandro Frank
and Roelof Bijker for invitation and kind hospitality.

\end{document}